\documentclass[prc,onecolumn,tightenlines,12pt]{revtex4}
\usepackage{graphicx}
\usepackage{dcolumn}
\usepackage{amsmath}
\newcommand{\be}{\begin{equation}}
\newcommand{\ee}{\end{equation}}
\newcommand{\bea}{\begin{eqnarray}}
\newcommand{\eea}{\end{eqnarray}}

\begin{document}

\author{J. X. de Carvalho$^{1,2}$, 
M. S. Hussein$^{\dag 1,2}$, 
M. P. Pato${^2}$ and A. J. Sargeant$^{2}$}
\affiliation{$^{1}$Max-Planck-Institut f\"ur Physik komplexer Systeme\\
N\"othnitzer Stra$\beta$e 38, D-01187 Dresden, Germany \\
$^{2}$Instituto de F\'{i}sica, Universidade de S\~{a}o Paulo\\
C.P. 66318, 05315-970 S\~{a}o Paulo, S.P., Brazil}
\title{Perturbative Treatment of Symmetry Breaking Within Random Matrix Theory
\thanks{Supported in part by the CNPq and FAPESP (Brazil).\\
        $^{\dag}$Martin Gutzwiller  Fellow, 2007/2008.}}

\begin{abstract}

We discuss the applicability, within the Random Matrix Theory, of perturbative treatment of symmetry breaking 
to the experimental data on the flip symmetry breaking in quartz crystal. We found that the values of the 
parameter that measures this breaking are different for the spacing distribution as compared to those 
for the spectral rigidity. We consider both twofold and threefold symmetries.
The latter was found to account better for the spectral rigidity than the former. Both cases, however,  underestimate
the experimental spectral rigidity at large L. This discrepancy can be resolved if an appropriate number of eigenfrequecies
is considered to be missing in the sample. Our findings are relevant to isospin violation study in nuclei.
\end{abstract}

\maketitle

The study of wave chaos using acoustic resonators \cite{Weaver},\cite{Tanner}
supplies an invaluable additional test of Random Matrix Theory (RMT) \cite{Meht, Meht1}. In a 1996 paper Ellegaard et al.\cite{Elleg}, studied the 
gradual breaking of the presumed twofold flip symmetry of a quartz crystal by removing an octant of a sphere
of an increasing radius at one of the corners and analysing the statistics of the resulting acoustic
eigenfrequencies. They found a gradual evolution of the spacing distribution from that of two uncoupled
Gaussian Othogonal Ensembles (2GOE) when the crystal is an uncut perfect rectangle, into a  single GOE, when a large
chunk of the crystal is removed from one of the corners of the rectangle. This constituted a complete breaking
of the symmetry present in the crystal in the uncut situation. The spectral rigidity, measured by Dyson's $\Delta_{3}(L)$ 
was also measured in this reference.  The 2 uncoupled GOE's was found to underestime by a great amount the large-$L$
data. This was attributed to pseudointegrable trajectories that do not suffer from the symmetry breaking. This point was
further analysed by \cite{Abul}. Using techniques developed by Pandey \cite{Pandey:1995}, Leitner \cite{Leitner} treated the symmetry
breaking problem with RMT-perturbation. He addressed only the spacing distribution. This work was further extended to 
the spectral rigidity in \cite{Abul-Magd:2004}. In all of the above treatment of the data of \cite{Elleg}, the 
assumption was made that the uncut crystal has a twofold flip
symmetry and thus is describable by two uncoupled GOE's. The treatment of Leitner \cite{Leitner} is found to describe fairly well
the NNL distribution, but fails for the spectral rigidity, in contrast to the exact numerical simulation
using the Deformed Gaussian Orthogonal Ensemble \cite{DGOE}, recently performed in \cite{last}.  In this paper we further analyse the perturbative treatment
of symmetry breaking within RMT. We find that the data of \cite{Elleg} can be accounted for with 3GOE's which are gradually
mixed till a 1GOE limit is attained. We further find that that if some levels were missing in the sample of eigenfrequecies
whose statistics is analysed, the $\Delta_{3}(L)$ can be very well accounted for even at large L  without the need for pseudointegrable trajectories,
whose calculation is difficult. 

Using appropriate perturbative methods Leitner \cite{Leitner} was
able to find a formula for 
the nearest neighbor distributio (NND) which contains the symmetry braking term. He started basically
with the formula for the nearest neigbhour spacing distribution
for the superposition of $m$ GOE's block matrices \cite{Meht}
\begin{equation}
  P_{m}(s) = \frac{d^{2}}{ds^{2}}E_{m}(s)
\end{equation}
where, for the case of all block marices having the same dimension one
has
\begin{equation}
  E_{m}(s) = \left( E_{1}(\frac{s}{m} ) \right)^{m},
\end{equation}
\begin{equation}
  E_{1}(x) = \int_{x}^{\infty}(1-F(t))\, dt,
\end{equation}
\begin{equation}
  F(t) = \int_{0}^{t}P_{1}(z)\,dz.
\end{equation}
In the above $P_{1}(z)$ is the normalized nearest neighbour spacing
distribution of one block matrix. It is easy to find for $P_{m}(s)$,
the following
\begin{eqnarray}
  P_{m}(s)& =& \frac{1}{m}\left[     \left(E_{1}(s/m)\right)^{m-1}
                                   P_{1}(s/m) + 
                             (m-1)(E_{1}(s/m))^{m-2}(1-F(s/m))^{2}
                        \right] \label{lago1} \\
          & \equiv & P^{(1)}_{m}(s) + P_{m}^{(2)}(s) \label{lago2}
\end{eqnarray}
If all the block matrices belong to the GOE, then one can use the
Wigner form for $P_{1}(z)$
\begin{equation}
  P_{1}(z) = \frac{\pi}{2}z e^{-\frac{\pi}{4}z^{2}}\approx \frac{\pi}{2}z,
\label{noiva1}
\end{equation}
thus
\begin{equation}
  F_{1}(z) = 1 - e^{-\frac{\pi}{4}z^{2}}\approx \frac{\pi}{4}z^{2},
\label{noiva2}
\end{equation}
\begin{equation}
  E_{1}(z) = erfc\left(\frac{\sqrt{\pi}}{2}z \right)\approx 1 - z.
\label{noiva3}
\end{equation}
where the large-$z$ limits of Eqs. (\ref{noiva1})-(\ref{noiva3}) are also
indicated above.
It is now clear that the above expression for $P_{m}(s)$, (\ref{lago1}) and
(\ref{lago2}), contains a term $P_{m}^{(1)}(s)$ with level repulsion,
indicating short-range correlation among levels pertaining to the
same block matrix and a second term $P_{m}^{(2)}(s)$ with no level
repulsion, implying short-range correlation among NND levels
pertaining to different blocks. Notice that for very small spacing, 
$P_{m}(s)$ behaves as 
\begin{equation}
  P_{m}(s) \approx \frac{\pi}{2m^{2}}s + \frac{m-1}{m}
\end{equation}
for $m=1$, we get the usual $P_{1}(0)=0$, while for $m>1$,
we get $P_{m}(0)=(m-1)/m$. 

To account for symmetry breaking,
Leitner \cite{Leitner} considered the mixing between levels pertaining
to nearest neigbhour block matrices and entails using the $2x2$ $P(s)$ 
distribution with full mixing. The DGOE result for the 2X2 matrix was derived in
 \cite{Carneiro:1991} and the resulting $P(s)$ is a product of a Poissonian term 
times a mixing term. Leitner's procedure \cite{Leitner} amounts to multiply
the factor $P_{m}^{(2)}(s)$ of Eq. (\ref{lago2}) by only the mixing term of the 2x2 $P(s)$ 
of \cite{Carneiro:1991} with the mixing parameter $\Lambda$
given by \cite{Pandey:1995}, $\Lambda = \lambda^2 {\rho^2}$, with $\lambda^2$ being the 
ratios of the variances of the matrix elements within a block matrix
to that of matrix elements pertaining to neighbouring off diagonal block matrices., and $\rho$ is the density of eigenfrequencies.
Thus, he found, assuming that $\Lambda << 1$,
\begin{equation}
  P_{m}(s,\Lambda)= P_{m}^{(1)}(s)+
                   P_{2\times 2}(s,\Lambda)P_{m}^{(2)}(s). \label{azuis}
\end{equation}

where $P_{2\times 2}(s,\Lambda)$ is given by \cite{Leitner}
\begin{equation}
P_{2\times 2}(s,\Lambda) = \sqrt{\frac{\pi}{8\Lambda}}I_{0}(\frac{s^{2}}{16\Lambda})
                    \exp\left(-\frac{s^{2}}{16\Lambda}\right),\label{sp}
\end{equation}

where $I_{0}$ is the modified Bessel function of order $0$.
Though $P_{m}(s)$ is normalized, $P_{m}(s,\Lambda)$ is not. Accordingly
one supplies coefficients $c_{N}$ and $c_{D}$, such that
\begin{equation}
  P_{m}(s,\Lambda,c_{N},c_{D})\equiv c_{N}P_{m}(c_{D}s,\Lambda)\label{verde}
\end{equation}
is normalized to unity. Similarly, $<s>$ should be unity too.
Eq. (\ref{azuis}) can certainly be generalized to consider the effect
of mixing of levels pertaining to next to nearest neighbour blocks,
and accordingly, $P_{3\times 3}(s,\Lambda)$, given in Ref. \cite{Carneiro:1991}
would be used
in Eq. (\ref{azuis}) instead of $P_{2\times 2}(s,\Lambda)$. In the 
following, however, we use Eqs. (\ref{azuis}), (\ref{verde}) as Leitner did \cite{Leitner}.

	In Ref. \cite{Leitner}, Leitner also obtained approximate expression 
for the spectral rigidity $\Delta_{3}(L)$ using results derived by 
French \emph{et al.}
\cite{French}. Leitner's approximation to $\Delta_{3}$ is equal to the GOE spectral 
rigidity plus perturbative terms, that is
\begin{eqnarray}
  \Delta_{3}^{(m)}(L;\Lambda)& \approx  & \Delta_{3}(L;\infty) +
        \frac{m-1}{\pi^{2}}\left[
	  \left(\frac{1}{2}-\frac{2}{\epsilon^{2}L^{2}}
                           -\frac{1}{2\epsilon^{4}L^{4}}\right) 
                           \right. \nonumber \\
        & & \times \ln (1+\epsilon^{2}L^{2}) +
            \frac{4}{\epsilon L}\tan^{-1}(\epsilon L ) +
           \left. \frac{1}{2\epsilon^{2}L^{2}} - \frac{9}{4} \right],
                                  \label{asa} 
\end{eqnarray}
where
\begin{eqnarray}
   \epsilon & = & \frac{\pi}{2(\tau + \pi^{2}\Lambda)}
\end{eqnarray}
For the cut off parameter we use the value \cite{Abul-Magd:2004}
$
\tau=c_me^{\pi/8-\gamma-1},
$
where $c_m=m^{m/(m-1)}$ 
and $\gamma\approx 0.5772$ is  Euler's constant.
This choice guarantees that when the symmetry is not
broken, $\Lambda = 0$, $\Delta_{3}^{(m)}(L,0)=m\Delta_{3}(L/m,\infty)$.
In Ref. \cite{Leitner:1997}, Leitner fitted 
Eq. (\ref{verde}) for $m$=2 to the NND from
Ref. \cite{Elleg}, however, he did not fit the spectral rigidity.
It is often the case that there are some missing levels in
the statistical sample analysed. Such a situation was
addressed recently by Bohigas and Pato \cite{ML}. These authors have
started from the general expression of $\Delta_{3}(L)$ derived by Dyson and Mehta \cite{Meht1}, namely,

\begin{equation}
\Delta_{3}(L) = \frac{L}{15} - \frac{1}{15L^{4}} \int_{0}^{L}~dx \big(L-x)^{3} (2L^{2} - 9xL-3x^{4}\big) Y_{2}(x), 
\end{equation}
where the two-point cluster function, $Y_{2}(x_{1},x_{2})$, which owing to translational invariance becomes a function
of the difference $x = \mid x_{1}-x_{2}\mid$, is defined by the usual expression,

\begin{equation}
Y_{2}(x_{1},x_{2}) = 1 - \frac{R_{2}(x_{1},x_{2})}{R_{1}(x_{1})R_{1}(x_{2})}, 
\end{equation}
where $R_{2}$ is the 2-point correlation function and $R_{1}$ is the density of the spectrum. 

If a fraction, $1-g$, of the levels were actually analysed , the cluster function remains invariant, apart from a rescaling of the relevant variables, when the unfolded spactrum is employed,  namely

\begin{equation}
Y_{2}^{g}(x_{1},x_{2}) = 1 - \frac{(1-g)^{2}R_{2}(x_{1}^{g},x_{2}^{g})}{(1-g)R_{1}(x_{1}^{g})(1-g)R_{1}(x_{2}^{g})} =
Y_{2} (x_{1}^{g},x_{2}^{g}), 
\end{equation}
where the scaled variables $x_{i}^{g}$ are just $\frac{x_{i}}{(1-g)}$

Using the above equation for the cluster function in the general expression for $\Delta_{3}(L)$, we obtain the
Missing-Level (ML) expression of \cite{ML}

\begin{equation}
  \Delta_{3}^{g}(L) = g\frac{L}{15}+(1-g)^{2}\Delta_{3}
                                \left(\frac{L}{1-g}\right).\label{dragao}
\end{equation}

In the application to our current problem of $m$-coupled GOE's, the above formula continue to be valid since
the basic input into its derivation, namely the invariance of $Y_{2}$, apart from the scaling of the argument $x$
into $x^{g}$, is quite general. Accordingly, we have the desired ML formula of $\Delta_{3}(L)$ for $m$-coupled GOE's,

\begin{equation}
\Delta_{3}^{(m)g}(L;\Lambda)= g\frac{L}{15}+(1-g)^{2}\Delta_{3}^{(m)}
                                \left(\frac{L}{1-g};\Lambda \right).\label{dragao}
\end{equation}

The presence of the linear term, even if small, could explain
the large $L$ behavior of the \emph{measured} $\Delta_{3}(L)$.
We call this effect the Missing Level (ML) effect. Another
possible deviation of $\Delta_{3}$ from Eq. (\ref{asa}) could arise
from the presence of pseudo-integrable effect (PI) \cite{Abul,bis2}.
This also modifies $\Delta_{3}$ by adding a Poisson term 
just like Eq. (\ref{dragao}).

The results of our analysis
are shown in figures 1 and 2. In Fig. 1, the 
sequence of six measured NNDs were fitted  for $m=2$ and $m=3.$ 
It can be seen that the Leitner model with three coupled GOE's give a comparable 
and in some cases even better fit than the $m=2$ one. Figure 1a in
fact shows a rather sharp peak in our calculated $P(s)$ for $m=3$,
$P_{3}(s,0.0056)$. We consider this a failure of the Leitner formula 
(\ref{verde})
for the uncut crystal. In fact, a more appropriate description of
the uncut crystal is to take $\Lambda = 0$, namely a superposition
of 3 uncoupled GOE's, which works almost as good as the 2 uncoupled GOE's
description. The other parts of figure 1, $(b)-(x)$ seem to show the
same insensitivity of $P_{m}(s,\Lambda)$ to $m$; the number of matrix blocks
used in DGOE description. It is this insensitivity of the short-range
nearest neighbour
level correlation, measured by the spacing distribution, to the assumed
symmetry inherent in the uncut crystal (and thus the number uncoupled
GOE's employed to describe it) that forces us to examine the long-range
level correlation, namely spectral rigidity, ``measured'' by Dyson's
$\Delta_{3}$ statistics.  

In Fig. 2. the 
$\Delta$-statistic was fitted with equation (\ref{asa}).
It is clear from the figure that a good fit to the data
of Ref. \cite{Elleg} is obtained with $m=3$ for the values
of $\Lambda$ given in table 1. This is to be contrasted with
the case of $m=2$  which, according to Eq. (\ref{asa}) results
in $\Delta_{3}^{(2)}(L,\Lambda )$ that is \emph{always} below
the one with $\Lambda=0$, $\Delta_{3}^{(2)}(L,0)$, which itself
is always below the data points of Ref. \cite{Elleg}. For this
reason, only the $\Delta_{3}^{(2)}(L,0)$ is shown in the figure.
It should be noted that the $\Delta$-statistics of the uncut crystal,
Fig. 2a is very well described by that of 3 uncoupled GOE's, 
namely $\Delta_{3}^{(3)}(L)=3\Delta_{3}^{(1)}(L/3)$ which is  
always larger than the above mentioned 
$\Delta_{3}^{(2)}(L)=2\Delta_{3}^{(1)}(L/2)$.
The most conspicuous exception is
 Fig. \ref{presente}b which corresponds to $r=0.5\, mm$
and  where $1414$ frequency eigenvalues were found. We consider
this a potential ML  case and take for $\Delta_{3}$, the
expression given in Eq. (\ref{dragao}) and use it in Eq.
(\ref{asa}) with $g$ taken as a parameter. The results are shown in Fig. 3. We find perfect fit to the \emph{data}, if $g$
is taken to be $6\%$, namely only $94\%$ of the eigenfrequencies
were in fact taken into account in the statistical analysis. In contrast, if 2GOE is used
we still do not get very good agreement even if 18\% of the levels are taken to be missing.
There is, threfore, room to account much better for all cases
(Fig. $2a$, $2c$, $\ldots$ ) in the 3GOE description, by appropiately choosing the 
correponding value of $g$. This is not the case if a 2GOE description is employed.

\begin{table}
\caption{\label{tab:inputs}Values of $\Lambda$ obtained by fitting
    Eqs. (\ref{verde}) and (\ref{asa}) respectively to the
    experimental NNDs and spectral rigidities from Ref. \cite{Elleg}.}
\begin{tabular}{|l|l|l|l|l|}
\hline
&\multicolumn{3}{c|}{$P(s)$}& $\Delta_3(L)$\\
\hline
 Data Set&  Ref. \cite{Leitner:1997}& Eq. (\ref{verde}) $m$=2 & Eq. (\ref{verde}) $m$=3 & Eq. (\ref{asa}) 
$m$=3\\
\hline
(a) & 0.0013 & 0.0030  & 0.0067  & 0.0056 \\
(b) & 0.0054 & 0.0063  & 0.0098  & 0.0016 \\
(c) & 0.0096 & 0.010  & 0.017    & 0.0017 \\
(d) & 0.0313 & 0.032 & 0.064     &  0.027 \\
(e) & 0.0720 & 0.070 & 0.13 & 0.050 \\
(f) & 0.113 & 0.12& 0.30 & 0.16 \\
(x) & 0.138 & 0.13 & 0.34 & 2.4 \\
\hline
\end{tabular}
\end{table}

In conclusion, the perturbative tratment of symmetry breaking within RMT is assessed
by using it to describe the coupling of $m$-fold symmetry 
The particular threefold case is used to analyse 
data on eigenfrequencies of elastomechanical vibration of a 
anisotropic quartz block. The treatment of Leitner \cite{Leitner} is found to describe fairly well
the NNL distribution, but fails for the spectral rigidity, in contrast to the exact numerical simulation
using the Deformed Gaussian Orthogonal Ensemble recently performed in \cite{last}. 
However, by properly taking into account the ML 
effect we have shown that the $\Delta_{3}(L)$, become closer to the data.

We have also
verified that if a 2GOE description is used, namely, $m=2$ , then an
account of the large-$L$ behaviour of $\Delta_{3}$ can also be obtained if a
much larger number of levels were missing in the sample. In our
particular case of Fig. 2b, we obtained $g = 0.18$. This is 3 times larger
than the ML needed in the 3GOE description. 
One major issue in the perturbative approach
is the need to use different sets of values of the mixing parameter $\Lambda$ for the
NNL distribution, $P(s)$, and for the $\Delta_{3}(L)$. Further study of this approach
is certainly required. Finally, we mention that the perturbation
approach to symmetry violation study within RMT of the type discussed in this paper may be valuable to isospin breaking in nuclei. In Ref. \cite{Mitch0} the breaking of isospin was studied
in the case of the nucleus $^{26}$Al where both T = 0 and 1 states
are present in the low lying spectrum, which required the use of 2GOE description. Other, heavier, odd-odd nuclei may exhibit a spectrum where T = 2 states may also be present, requiring a 3GOE description of the isospin symmetry breaking. Work along these lines is in progress.

\newpage

\begin{figure}[h]
\includegraphics[angle=270,width=\textwidth]{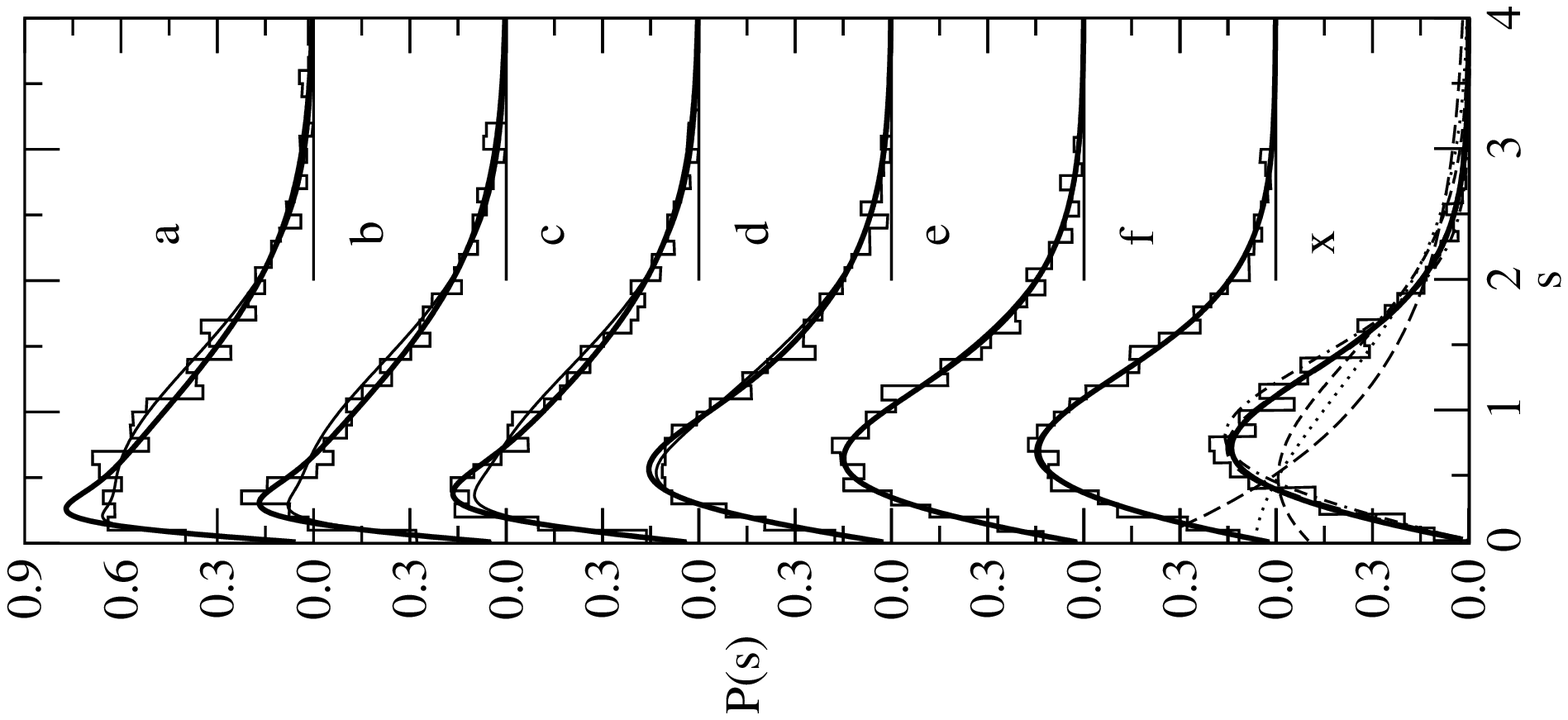}
\caption{
Nearest Neighbour Distributions. Histograms show data (a)-(x) from
Ref. \cite{Elleg}. Thin and thick solid lines show fits to the data
carried out using Eq. (\ref{verde}) with m=2 and m=3 respectively. In
graph (x) the long-dashed line is the Poisson distribution, the dot-dashed
line is the Wigner distribution and the dashed and dotted lines are
the respective distributions for superpositions of 2 and 3 uncoupled GOEs.
See Table I for the values of $\Lambda$ obtained from the fits and the
text for details.
} \label{eterno}
\end{figure}
\begin{figure}
\includegraphics[angle=270,width=\textwidth]{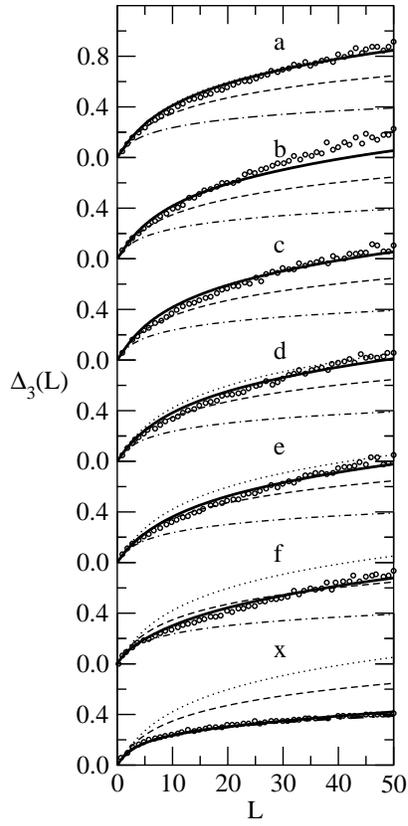}
\caption{
Spectral Rigidities. Circles show data (a)-(x) from
Ref. \cite{Elleg}. Thick solid lines show fits to the data
carried out using Eq. (\ref{verde}) with m=3. The dot-dashed
line is the GOE spectral rigidity and the dashed and dotted lines are
the respective rigidities for superpositions of 2 and 3 uncoupled GOEs.
See Table I for the values of $\Lambda$ obtained from the fits.
} \label{presente}
\end{figure}
\begin{figure}
\includegraphics[width=\textwidth]{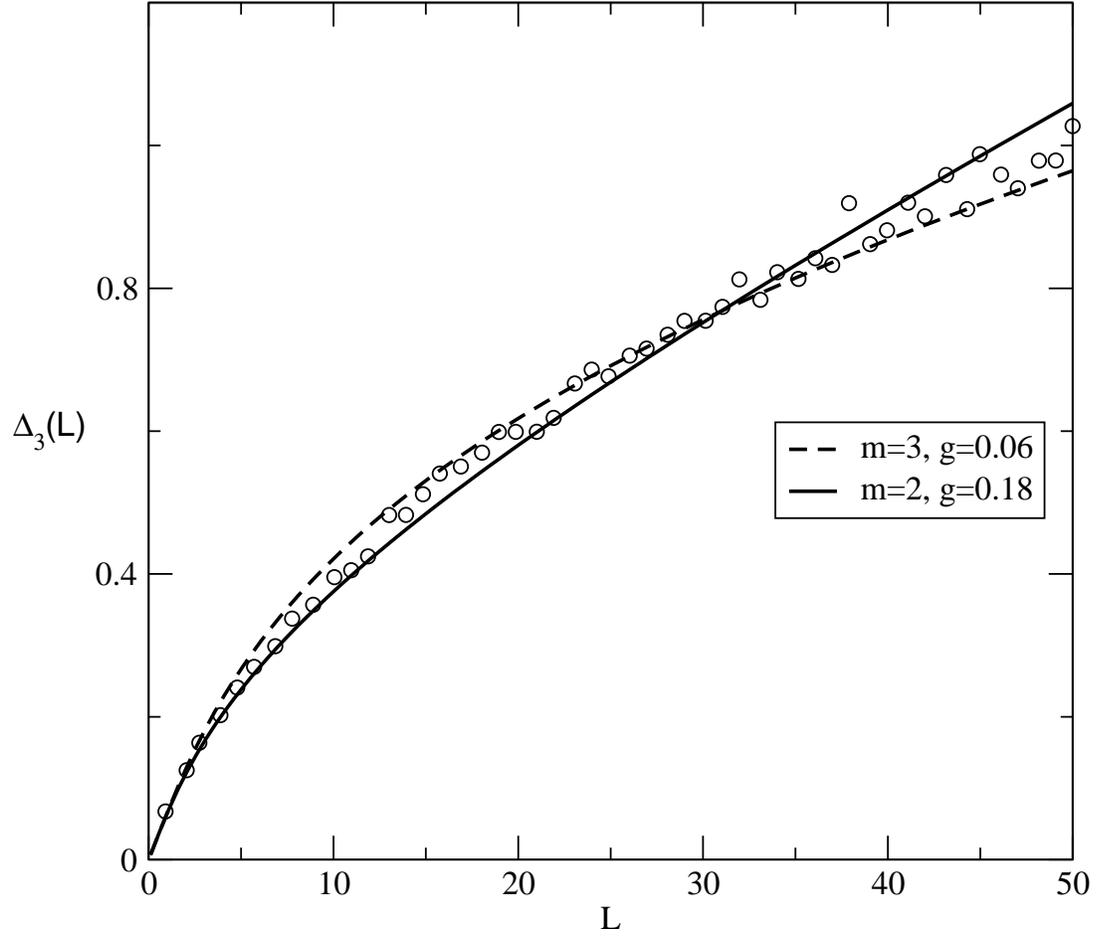}
\caption{
Missing Levels. Full line corresponds to $\Delta_{3}(L)$ for the 3 GOE's of Fig 2b with g = 0.06, while 
the dashed line corresponds to to the 2 GOE's with g = 0.18. The data points corresponding to the case with r = 0.5 mm,
are from fig 2b of \cite{Elleg}  
} \label{futuro}
\end{figure}

\end{document}